\newlength{\extralineskip}

\documentstyle[draft]{article}
\makeatletter
\newdimen\normalarrayskip              
\newdimen\minarrayskip                 
\normalarrayskip\baselineskip
\minarrayskip\jot
\newif\ifold             \oldtrue            \def\new{\oldfalse}
\def\arraymode{\ifold\relax\else\displaystyle\fi} 
\def\eqnumphantom{\phantom{(\theequation)}}     
\def\@arrayskip{\ifold\baselineskip\z@\lineskip\z@
     \else
     \baselineskip\minarrayskip\lineskip2\minarrayskip\fi}
\def\@arrayclassz{\ifcase \@lastchclass \@acolampacol \or
\@ampacol \or \or \or \@addamp \or
   \@acolampacol \or \@firstampfalse \@acol \fi
\edef\@preamble{\@preamble
  \ifcase \@chnum
     \hfil$\relax\arraymode\@sharp$\hfil
     \or $\relax\arraymode\@sharp$\hfil
     \or \hfil$\relax\arraymode\@sharp$\fi}}
\def\@array[#1]#2{\setbox\@arstrutbox=\hbox{\vrule
     height\arraystretch \ht\strutbox
     depth\arraystretch \dp\strutbox
     width\z@}\@mkpream{#2}\edef\@preamble{\halign \noexpand\@halignto
\bgroup \tabskip\z@ \@arstrut \@preamble \tabskip\z@ \cr}%
\let\@startpbox\@@startpbox \let\@endpbox\@@endpbox
  \if #1t\vtop \else \if#1b\vbox \else \vcenter \fi\fi
  \bgroup \let\par\relax
  \let\@sharp##\let\protect\relax
  \@arrayskip\@preamble}
\def\eqnarray{\stepcounter{equation}%
              \let\@currentlabel=\theequation
              \global\@eqnswtrue
              \global\@eqcnt\z@
              \tabskip\@centering
              \let\\=\@eqncr
              $$%
 \halign to \displaywidth\bgroup
    \eqnumphantom\@eqnsel\hskip\@centering
    $\displaystyle \tabskip\z@ {##}$%
    &\global\@eqcnt\@ne \hskip 2\arraycolsep
         $\displaystyle\arraymode{##}$\hfil
    &\global\@eqcnt\tw@ \hskip 2\arraycolsep
         $\displaystyle\tabskip\z@{##}$\hfil
         \tabskip\@centering
    &{##}\tabskip\z@\cr}
\makeatother
\def\tr#1{{\rm tr}\kern-3pt\left[#1\right]}

\def\nn{\nonumber}

\def\beq{\begin{equation}}
\def\eeq{\end{equation}}
\def\be{\beq\new\begin{array}{c}}
\def\ee{\end{array}\eeq}

\newcommand{\sect}[1]{\setcounter{equation}{0}\section{#1}}
\renewcommand{\theequation}{\thesection.\arabic{equation}}

\input epsf

\begin{document}

\begin{titlepage}
\setcounter{footnote}0
\begin{center}

\phantom . \hfill ITEP-M7/95 \\
\phantom . \hfill OU-HET-232 \\
\vspace{0.3in}

{\Large{INTEGRABILITY and SEIBERG-WITTEN THEORY}}
\\[.2in]
{\it H.Itoyama\footnote{Departement of Physics, Graduate School of Science,
Osaka University, Toyonaka, Osaka 560, Japan. 
E-mail address: itoyama@funpth.phys.sci.osaka-u.ac.jp} and 
A.Morozov\footnote{117259, ITEP, Moscow,  Russia. 
E-mail address: morozov@vxitep.itep.ru}}
\\[.2in]

{\it Contribution to the Proceedings of the Conference}\\
{\it ``Frontiers in Quantum Field Theory''}\\
{\it in honor of Professor Keiji Kikkawa's 60th birthday }\\

\end{center}
\bigskip
\bigskip

\centerline{\bf ABSTRACT}
\begin{quotation}
A summary of results is presented,  which provide exact description
  of the low-energy 
$4d$ $N=2$ and $N=4$ SUSY gauge theories in terms of $1d$
integrable systems.

\end{quotation}
\end{titlepage}

\setcounter{footnote}{0}

This conference is devoted to the impact that 
the ideas of Professor K.Kikkawa -
especially those concerning duality and string field theory -
have  had in  modern theoretical physics  as well as to their interplay with
the other concepts and methods. 
The subject of this contribution is related to  duality
through  the remarkable achievement of N.Sei\-berg and E.Witten
\cite{SW,SW2}, who used  the duality to obtain the explicit answers for the
low-energy effective actions of certain four-dimensional gauge theories.
Below is a brief presentation  of the results of papers \cite{Go,IM,IM2},
which - as a minimum - allow to represent the answers of \cite{SW}
and \cite{SW2} in a simple and compact form, and - as a maximum -
should attract attention to the pertinent role that integrability (and thus
abstract group theory) plays in the description of exact effective actions,
 namely, the role which becomes especially pronounced in the low-energy
(Bogolubov-\-Whitham or topological) limit of quantum field theory.
See refs.\cite{MW1}-\cite{GHL} for related developments.
All the relevant references can be found in \cite{IM,IM2}.

The presentations of K.Intriligator and T.Eguchi at this conference
allow us not to repeat all the discussion of $4d$ physics and the
methods used in \cite{SW,SW2}. We can directly proceed to our main
subject.

\sect{The problem: from $4d$ to group theory}

Renormalization group (RG) flow for the $4d$ $N=2$ SUSY YM theory is
schematically shown in Fig.1.
$${\epsfxsize=8 truecm \epsfbox{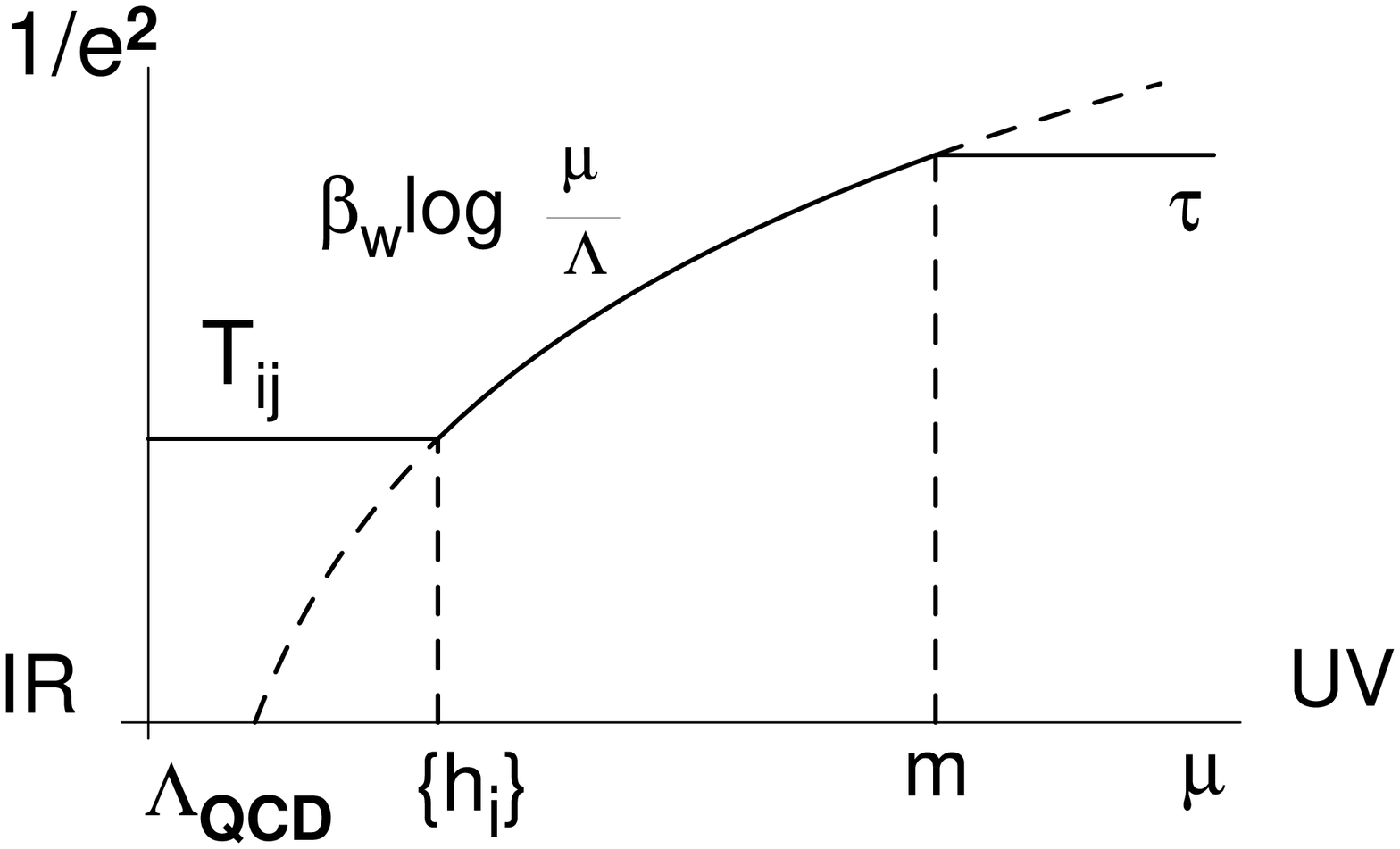}}$$
Perturbative expression 
$\frac{1}{e^2(\mu)} \sim \beta_W\log\frac{\mu}{\Lambda}$
receives, of course, nonperturbative corrections, which do not change
the shape of the curve too much, except at the infrared (IR). 
We assume that  theory is {\it softly} regularized in the 
ultraviolet (UV) by being embedded into some 
UV-finite theory, say, $N=4$ SUSY YM. ( An alternative possibility
could be the $N_f = 2N_c$ model). Such regularization affects the
theory for $\mu > m$. From the UV perspective $m$
introduces the mass scale in conformally invariant theory at 
$\mu \gg m$.\footnote{
Actually, $m$ can be identified with the mass of the extra 
$N=2$ SUSY supermultiplet in the adjoint of the gauge group, 
which is added to the pure  $N=2$
SUSY YM in order to convert it into the $N=4$ SUSY one in the UV.}

In the IR the model is {\it softly} regularized by   the condensates of the
scalar fields, $h_k = \frac{1}{k}\langle {\rm tr}\ \phi^k\rangle$,
which break the gauge symmetry down to the pure abelian one.
Then all the {\it minimal} couplings of fields (which are in adjoint of
the gauge group) disappear.
In the UV the $N=4$ SUSY YM model is completely
characterized by the gauge group $G$ and the bare coupling
constant $\tau = \frac{4\pi i}{e^2} + \frac{\vartheta}{2\pi}$
(a single complex number). The abelian model in the IR contains
$r_G = {\rm rank}\ G\ $ abelian gauge fields, and the corresponding
complex charges form a matrix $T_{ij}$, $\ i,j=1,\ldots,r_G$:
\be
\begin{array}{ccc}
S^{\rm (IR)} = \int d^4x \sum_{i,j=1}^{r_G} {\rm Im}\
T_{ij} (G^iG^j + iG^i\tilde G^j) + \ldots  & \ \ & 
S^{\rm (UV)} = \int d^4x\ {\rm Im}\ \tau {\rm tr}(G^2 + G\tilde G)
+ \ldots  \\
=\int d^4xd^2\theta \sum_{i,j=1}^{r_G} {\rm Im}\ 
T_{ij}{\cal W}^i{\cal W}^j + \ldots~  &  & ~
= \int d^4xd^2\theta\ {\rm Im} \tau {\rm tr} \ W^2 + \ldots   \\
=\int d^4xd^4\theta \ {\rm Im}\ {\cal F}(\Phi^i) & &
=\int d^4xd^4\theta \ {\rm Im}\ F(\Phi)~~~~.
\end{array}
\label{IRUVS}
\ee
The three lines here are written in the $N=0$, $N=1$ and $N=2$ 
notation respectively. Only  kinetic terms for the gauge (super)fields
are presented explicitly and  the rest is denoted by dots. The $N=2$
(and $N=4$) superfields in the UV are non-abelian and thus contain
infinitely many auxiliary fields. At the same time the abelian 
superfields in the IR are equaly simple for $N=1$ and $N=2$ SUSY:
\be
\Phi ^i = A^i(\theta) + \hat\theta{\cal W}^i(\theta) + \ldots =
a^i + \theta\hat\theta G^i + \ldots\;.
\ee
As a result, the abelian charge matrix is just
\be
T_{ij}(a) = \frac{\partial^2{\cal F}}{\partial a^i\partial a^j}.
\ee
Duality properties become transparent when one introduces a
''dual scalar'':
\be
a_i^D = \frac{\partial {\cal F}}{\partial a^i}, \ \ \ {\rm so\ that}\ \ \ 
T_{ij} = \frac{\partial a_i^D}{\partial a^j}.
\ee
It is a distinguished property of $N=2$ gauge models that the 
Wilsonian background fields - of which $a^i$ and $a_i^D$ are
examples - and not just the moduli $h_i,\ \tau,\ m$ - are
physical observables. This is due to the important result of
\cite{OW}, which says that the mass spectrum
(of BPS saturated states) - a physical observable - is
exactly given by ${\cal M} \sim | {\displaystyle
 \sum_i(m_ia^i + n^ia_i^D) }|$. 

The problem of defining the low-energy effective action in this
framework can be formulated as follows:
INPUT:  $\ G$ (the gauge group), $\tau$ (the UV bare coupling constant),
$m$ (mass scale) and $h_i$ (symmetry breaking v.e.v.'s).
OUTPUT: $\ a^i(h)$ (the background fields) and ${\cal F}(a^i)$ 
(the prepotential)
- and thus also $a_i^D = \partial{\cal F}/\partial a^i$ and $T_{ij}(a) =
\partial^2{\cal F}/\partial a^i\partial a^j$.
In other words, all what one seeks for 
in the setting of Seiberg-Witten theory is the RG map:
\be
\left\{G,\ \tau;\ m,\ h_i\right\} \ \longrightarrow \
\left\{ a^i(h);\ {\cal F}(a^i)\right\}
\label{map}
\ee
One can see here an analogy with the $c$-theorem of A.Zamolodchikov
in $d=2$ physics: the detailed description of  RG flow from one fixed
point (conformal model) to another is somewhat sophisticated, but the net
result is simple: when we get from the UV to the IR it is enough to say
that the central charge  has  jumped to its
 adjacent value (in the corresponding
series of conformal models, specified by the symmetries preserved
during the RG flow).

\sect{Intermediate data: Riemann surfaces}

The discovery of N.Seiberg and E.Witten was that the mapping (\ref{map})
is actually decomposed into two steps:
\be
\left\{ G,\ \tau;\ m,\ h_i\right\} \ \longrightarrow \
\left\{\begin{array}{c}
{\rm Riemann\ surface}\ {\cal C}; \\
{\rm meromorphic\ one-differential}\ d{S}_{\rm min}\ {\rm on}\ {\cal C} \\
{\rm with\ the\ property}\ \frac{\partial dS_{\rm min}}{\partial h_i} =
{\rm holomorphic}
\end{array}\right\}; \nn \\
\left\{{\cal C},\ dS_{\rm min}\right\} \ \longrightarrow \
 \left\{a^i(h);\ {\cal F}(a)\right\}
\label{map'}
\ee

The second  step is simple: given ${\cal C}$ one can define
the set of conjugate $A$ and $B$ cycles on it, and given $dS_{\rm min}$
one can write
\be
a^i(h) = \oint_{A_i} dS_{\rm min}, \ \ \ 
a_i^D(h) = \oint_{B^i} dS_{\rm min}\;, 
{\cal F}(a) = \left.\frac{1}{2}\sum_I \oint _{A_I}dS_{\rm min} \int_{B^I}
dS_{\rm min} \right|_{h_i = h_i(a_j) } \;\;.
\label{pre}
\ee
In the last formula (the one for the prepotantial), the set of
$A$ and $B$ contours should be enlarged to include those wrapping
around and connecting the singularities of  $dS_{\rm min}$: see
\cite{IM2} and section \ref{prepo} below. Also, this formula,
 when $dS_{\rm min}(h)$ is substituted into , gives an answer as 
a function of $h_i$. One should further express $h_i$ by $a^i$
 with the help of   the first formula  and substitute it into the last one in
order to obtain the prepotential ${\cal F}(a^i)$.  

The genus of Riemann surface ${\cal C}$  need not coincide
with $r_G = {\rm rank}\ G$: it is often larger. Then eqs.(\ref{pre}) 
 seem senseless since the number of $A$ and $B$ contours 
on the r.h.s. can exceed that of the $a^i$ and $a_i^D$ on the l.h.s.
However, in such cases - given appropriate $dS_{\rm min}$ 
(as defined in eq.(\ref{dSmin}) below) -
all the extra integrals at the r.h.s. of (\ref{pre}) automatically vanish
- and there are exactly $r_G$ non-vanishing $a^i$ - as necessary.
This is of course also important to make the dependence $a^i(h)$
invertible.
Thus the second mapping in (\ref{map'}) is completely described 
and very simple. The real issue is the first mapping in (\ref{map'}).

\sect{What is ${\cal C}$?  ~emergence of integrable systems}

This mapping
\be
\left\{G,\ \tau;\ m,\ h_i\right\} \ \longrightarrow \
\left\{{\cal C};\ dS_{\rm min}\right\}
\label{gra}
\ee
contains no reference to four dimensions, Yang-Mills theory
or anything like that: it is clearly something much simpler 
and general. With no surprise, it can be described in a language
far more primitive than that of $4d$ gauge theories: one should
look for  this mapping at the first place where the group theory
 (the input in (\ref{gra})) 
meets with the algebraic geometry (the output of (\ref{gra})).
A natural place of such kind is integrability theory \cite{Kr}.
 Namely, the
map (\ref{gra}) possesses description in terms of $1d$ integrable
models.
In other words, the particular question (the $UV\ \longrightarrow\ IR$
map) in $4d$ gauge theory appears to be equivalent to some
(actually, almost the same) question in the $1d$ integrability theory.

In this language theory  (\ref{gra}) can be described
as follows:
Given a simple Lie group $G$ one can construct an associated $1d$
integrable model. Parameters $\tau$ and $m$ naturally appear in
this construction. The only thing that we  need on the
 model emerged is its Lax operator $L(z)$, which is a $G^*$-valued
function (matrix) on the phase space of the system and depends 
also on the ''spectral parameter'' $z$. Thus,\footnote{
The map (\ref{grouptoLax}) is actually a canonical one in the framework of
geometrical quantization (Kirillov-Kostant method): it can be
nicely described in terms of coadjoint orbits of $G$, $G_q$ and
$\hat G$, momentum maps, Hitchin varieties etc. What we need
to know here about this map is much simpler: 
that it exists, is naturally defined 
entirely in terms of group theory, and most of explicit formulas
(in convenient coordinates) are well-known since 1970's.
}
\be
G\ \stackrel{\tau, m}{\longrightarrow}\ L(z).
\label{grouptoLax}
\ee
Then the first ingredient of the map (\ref{gra}) is
\be
{\cal C}:\ \ \ \ \det \left( t - L(z) \right) = 0.
\label{specu}
\ee
If the spectral parameter $z$ in (\ref{grouptoLax}) belongs to some complex
''bare spectral surface'' $E$, this equation defines the spectral curve 
${\cal C}$  of  $L(z)$ as a ramified covering over $E$. (For every
$z \in E$ there are several points on ${\cal C}$, differing by the
choice of the eigenvalue $t$ of $L(z)$. The sheets are glued together
at the points where some eigenvalues coincide.)
For the given $L(z)$ eq.(\ref{specu}) depends on the integrals
of motion of integrable system. These are identified with the moduli 
$h_i$ at the l.h.s. of (\ref{map}).

\sect{Examples}
\subsection{$1d$ sine-Gordon model ($G = SL(2)$) \label{sG}}
In this case
\be
L(z) = \left(
\begin{array}{cc}  p & 1 + \frac{1}{z}e^{-q} \\ 
         e^q + z & -p 
\end{array}\right) \;\;.
\ee
The Lax operator depends on the coordinate $q$ and the momentum $p$,
but the spectral curve (\ref{specu}) depends only on their
particular combination $h_2 = p^2 + (e^q+e^{-q}) = p^2 + 2\cosh q$:
\be
(t - p)(t+p) = e^q + e^{-q} + z + \frac{1}{z},
\ee
or
\be
z + \frac{1}{z} = t^2 - h_2.
\ee
This $h_2$ is nothing but the integral of motion (the second Hamiltonian)
of the sine-Gordon system. Thus we see that what remains in (\ref{specu}) 
from the phase-space dependence of the Lax operator is just 
that on the constants of motion (this is one of the central
facts in integrability theory). For us this means, that once $h_i$ are
identified with the integrals of motion of integrable system, we indeed
get a map $\{G, \tau, m, h_i \} \rightarrow {\cal C}$.

\subsection{$SL(2)$ $1d$ Calogero system \label{Casec}}
This time the Lax operator is expressed through elliptic functions.
Elliptic functions live on elliptic {\it bare} spectral curve $E(\tau)$.
Its modulus $\tau$ is exactly the one which is identified with the
{\it bare} coupling constant in $4d$ theory.
One can choose a coordinate on $E(\tau)$ in two essentially different
ways:  the one is  the flat coordinate $\xi$ and elliptic functions are
(quasi) doubly periodic in $\xi$  and the other is  the
 elliptic parametrization
\be
E(\tau): \ \ \ y^2 = (x - \hat e_1(\tau))
(x - \hat e_2(\tau))(x - \hat e_3(\tau)),
\ee
and $\hat e_1(\tau) - \hat e_2(\tau) = \theta_{00}^4(\tau)$, etc.
The Lax operator of Calogero-Moser model is \cite{KriCaLax} 
\be
L(\xi) = \left(\begin{array}{cc}
p & gF(q|\xi) \\
gF(-q|\xi) & -p \end{array}\right)
\ee
where $F(q|\xi) = \frac{\sigma(q+\xi)}{\sigma(\xi)\sigma(q)}$
and Weierstrass function
\be
x \sim \wp(\xi) = -\partial^2_\xi \log\sigma(\xi) =
\frac{1}{\xi^2} + {\sum_{m,n}}' \left(\frac{1}{(\xi + m + n\tau)^2}
- \frac{1}{(m + n\tau)^2}\right).
\ee
The full spectral curve is now
\be
{\cal C}: \ \ \ \ \det \left( t - L(\xi)\right) = 0,  \nn \\
(t - p)(t + p) = g^2F(q|\xi)F(-q|\xi) = g^2\left(\wp(\xi) - \wp(q)\right),
\ee
or simply
\be
{\cal C}: \ \ \ \  g^2\wp(\xi) = t^2 - h_2,
\ee
where this time $h_2 = p^2 + g^2\wp(q)$ is the second Hamiltonian
of Calogero-Moser system.
Calogero coupling constant $g$ is to be identified with  the parameter $m$
in $4d$ considerations:
\be
g^2 \sim m^2.
\ee

Thus, in the framework of Calogero-Moser models we devised a mapping
(\ref{grouptoLax}) parametrized by two variables, $\tau$ and $m \sim g$,
  which is exactly what necessary for our purposes.

\sect{What is $dS_{\rm min}$?}

The last thing which is necessary to formulate our description 
of Seiberg-Witten theory is  an explicit expression for 
$dS_{\rm min}$, which enters the formulas (\ref{pre}) for the background
fields $a^i$ and $a_i^D$.
Now we know that the appropriate {\it bare} spectral
surface is the elliptic curve $E(\tau)$ and the full spectral curve
${\cal C}$ is a ramified covering over $E(\tau)$ defined by the
equation $\det (t - L(\xi)) = 0$.  We are ready
to give an explicit expression for $dS_{\rm min}$.
On $E(\tau)$ there is a distinguished canonical holomorphic
1-differential  
\be
d\omega_0 = d\xi = \frac{1}{2\pi}\frac{dx}{y(x)} = 
\frac{1}{2\pi}\frac{dx}{\sqrt{x-\hat e_1(\tau))(x - \hat e_2(\tau))
(x - \hat e_3(\tau))}}
\ee   
Its periods on $E(\tau)$ are $1$ and $\tau$.
$dS_{\rm min}$ is just twice the product of the Lax-operator eigenvalue $t$
and this $d\omega_0$:\footnote{
Again, as in the case of  (\ref{grouptoLax}),
there are different interpretations of this formula: their origins
range from the theory of prepotential (quasiclassical $\tau$-functions)
and Bogolubov-Whitham theory to Hamiltonian structures of
integrable theories and symplectic geometry of Hitchin varieties.
Again, the only thing that is essential for us is that all these
interpretations are essentially the same and that  the explicit formula
(\ref{dSmin}) is true. }
\be
dS_{\rm min} = 2td\omega_0.
\label{dSmin}
\ee

\sect{Examples}
\subsection{$SL(2)$ Calogero and its limiting cases}

Since the curve ${\cal C}$ in this case is given by eq.
\be
t^2 - h = g^2\wp(\xi),
\ee
eq.(\ref{dSmin}) says that
\be
dS_{\rm min} \sim \sqrt{h + g^2\wp(\xi)}\ d\xi 
\ \stackrel{x \sim \wp(\xi)}{\sim}\ 
\frac{\sqrt{h - g^2x}}{y(x)}dx.
\label{dScal}
\ee
Despite the full spectral curve ${\cal C}$ is of genus 2
(${\cal C}$ is obtained by gluing two copies of $E(\tau)$ along
one cut, which connects two points with  $x = g^{-2}h$ and
two different values of 
$y = \pm\sqrt{\prod_{a=1}^3 (\frac{h}{g^2} - \hat e_a)}$).
However, the differential (\ref{dScal}) can be essentially considered
as living on some other - genus {\it one}  curve
\be
\hat{\cal C}: \ \ \ \ \hat y^2(x) = (h - g^2x)
(x - \hat e_1(\tau))(x - \hat e_2(\tau))(x - \hat e_3(\tau)),
\ee
which is different from $E(\tau)$ (except for the $N=4$ SUSY limit
of $h = \infty$), but of which ${\cal C}$ is also a double covering.
In fact, this is another way of saying that $dS_{\rm min}$ has only
two - rather than four (= twice the genus of ${\cal C}$) -
non-vanishing periods: this is an example  of how the extra periods
are automatically eliminated by the choice of peculiar 1-differential
(\ref{dSmin}).
In terms of $\hat{\cal C}$ we have
\be
dS_{\rm min} \sim \frac{h - g^2x}{\hat y(x)} dx,
\ee
and we remind once again that $g \sim m$ = the mass of adjoint
matter multiplet.

There are two interesting limits of this formula that deserve attention.
The first one is the $N=4$ SUSY limit, when the adjoint multiplet gets
massless: $m^2 \sim g^2 = 0$. Then obviously 
\be
\hat{\cal C} \rightarrow E(\tau), \nn \\
dS_{\rm min} \rightarrow 2\sqrt{h}d\omega_0,
\ee
and the periods (background fields) are
\be
a = \oint_A dS_{\rm min} \rightarrow 2\sqrt{h}, \ \ \ \
a^D = \oint_B dS_{\rm min} \rightarrow 2\tau\sqrt{h}.
\ee
Of somewhat more interest is the opposite limit, when the
matter multiplet decouples, $m \rightarrow \infty$. Of physical
interest is, however, the situation when the mass scale survives,
 i.e.  the case  that the dimensional transmutation takes place.
This is achieved in the double scaling limit, when simultaneously
\be
m^2 \sim g^2 \rightarrow \infty, \nn \\
\tau \rightarrow i\infty, \ \ \ {\rm or} \ \ \
 q \equiv e^{i\pi\tau} \rightarrow 0,
\ee
but 
\be
\Lambda^{N_c} = m^{N_c}q, \ \ \ {\rm i.e.}\ \ \   \Lambda^2 = m^2q
\ee
remains finite. 
In this limit the {\it bare} spectral curve $E(\tau)$ (parametrized by $\xi$) 
degenerates into
a double punctured sphere with coordinate $z$, and
\be
\xi \rightarrow \frac{1}{2\pi i}\log \frac{z}{q}, \ \ \ {\rm and} \ \ \
d\omega_0 = d\xi \rightarrow \frac{1}{2\pi i}\frac{dz}{z}.
\ee
Since $g^2 \sim \frac{1}{q}$ and $\xi + n\tau \sim \frac{1}{2\pi i}\log
(zq^{2n-1})$, it is clear that the only two terms which survive in the sum
\be
g^2x \sim g^2\wp(\xi) \sim \sum_{n} \frac{g^2}{\sinh^2 \pi(\xi + n\tau)} -
{\sum_n}'\frac{g^2}{\sinh^2\pi n\tau}
\ee
in the double scaling limit are those with $n = 0$ and $n=1$, so that
\be
x \sim \wp(\xi) = -\left(C(\tau) + 4q(z + z^{-1}) + o(q^2)\right).
\ee
Here
\be
C(\tau) = \frac{1}{3\pi i}\frac{\partial\log \Delta(\tau)}{\partial \tau},
\ \ \ \ \Delta(\tau) = q^2\prod_{n=1}^\infty (1 - q^{2n})^{24}.
\label{Cdef}
\ee 
Accordingly, the scaling rule for the $h$ parameter is
\be
h + g^2C(\tau) = \frac{1}{2}u,
\ee
and it is $u$ that remains finite in the double sacling limit (while
$h \rightarrow \infty$ as $-g^2C(\tau = \infty) \sim -\frac{1}{3q}$).
As a result,
\be
dS_{\rm min}^{Cal} = 2\sqrt{h + g^2\wp(\xi)} d\xi  
\ \stackrel{d.s.l.}{\longrightarrow} \ 
\frac{1}{\pi\sqrt{2}}\sqrt{\frac{\tilde x - u}{\tilde x^4 - \Lambda^4}}
d\tilde x \ \sim \ \sqrt{u - \Lambda^2\cos\varphi}\ d\varphi,
\label{N=2ldS}
\ee
where $\tilde x = -\frac{1}{2}\Lambda^2(z+ z^{-1}) = -\Lambda^2
\cos\varphi$. The r.h.s. of (\ref{N=2ldS}) is exactly the original
Seiberg-Witten differential of \cite{SW}, which
describes the $N=2$ SUSY pure gauge
$SL(2)$  model. It can be of course immediately reproduced from
the sine-Gordon description of our section \ref{sG}. See \cite{IM2} for 
more details.

\subsection{Toda chain for any $SL(N_c)$ and beyond}

In this case the {\it bare} spectral surface is a double-punctured
sphere obtained by degeneration of elliptic $E(\tau)$.
Other formulas from s.\ref{sG} are generalized as follows:
\be
L(z) = {\vec p}{\vec H} + 
\sum_{{\rm simple}\  {\vec \alpha} > {\vec 0}} 
\left(E_{{\vec \alpha}}  + 
e^{{\vec \alpha}{\vec q}}E_{-{\vec \alpha}}\right) +
zE_{{\vec \alpha}_0} + \frac{1}{z} e^{{\vec \alpha}_0{\vec q}}
E_{-{\vec \alpha}_0}.
\label{TCLax}
\ee
In the fundamental representation of $GL(N_c)$ the roots are represented 
as matrices $E_{ij}$ with  non-vanishing entries at the crossing
of $i$-th row and $j$-th column.  For positive roots $i <j$ (upper triangular
matrices), for negative roots $i>j$. Diagonal matrices represent Cartan
elements. The simple positive/negative roots belong to the first
upper/lower subdiagonal, the affine roots $\pm {\vec \alpha}_0$ are
located at the left lower/ right upper corner respectively. Thus
$$
L(z) = \left(\begin{array}{cccccc}
p_1  & 1      &      0  &       &  0  &  \frac{1}{z}e^{q_1-q_{N_c}} \\
e^{q_2-q_1}      & p_2  &      1  &       &  0  &  0  \\
0       &   e^{q_3 - q_2}  &   p_3  & \ldots  &  0 & 0 \\
 & & \ldots & & &  \\
0 & 0 & 0 &  & p_{N_c-1} & 1 \\
z & 0 & 0 & & e^{q_{N_c}-q_{N_c-1}}& p_{N_c}
\end{array} \right) .
$$
The full spectral curve is given by the equation
\be
0 = \det \left(t - L(z)\right) = \nn \\
= (t-p_1)(t-p_2)\ldots(t-p_{N_c}) + \sum_{i=1}^{N_c} e^{q_{i+1} - q_i} 
+ \ldots - z - \frac{1}{z} = \nn \\
= \sum_{l=0}^{N_c} S_l(h_k)t^{N_c - l} = 2P_{N_c}(t|h).
\label{TCcur}
\ee
In this formula one should take into account the periodicity condition
$q_{N_c+1} = q_1$. As usual all the $p,q$-dependent terms gather
into the Toda Hamiltonians $h_k$, $\ h_1 = \sum_i p_i$, $\ h_2 = 
\sum_i\left(\frac{1}{2}p_i^2 + e^{q_{i+1} - q_i}\right)$,
 $\ldots$ ($h_1 = 0$ for 
$G = SL(N_c)$). Finally, $S_l(h)$ are Schur polynomials. One can
easily see how they appear by omitting all the interaction terms
(with $q$-exponents) and substituting the free Hamiltonians
$h_k^{(0)} = \frac{1}{k}\sum_i p_i^k$ for $h_k$. In order to introduce
  the interaction back it is enough to substitute back $h^{(0)}_k \rightarrow
h_k$: all interaction effects enter only through $h_k$.
Thus we obtain the spectral curve in the form:
\be
{\cal C}: \ \ \ \ z + \frac{1}{z} = 2P_{N_c}(t).
\label{TCcurve}
\ee
It can be brought into a more familiar form $Y^2 = P_{N_c}^2(t) - 1$
by a substitution $2Y = z - z^{-1}$.
If $N_f$ $N=2$ SUSY matter superfields 
in the fundamental representation of the gauge group
are added to the model in $4d$, the curve (\ref{TCcurve}) is replaced by
\be
z + \frac{1}{z} = \frac{2P_{N_c}(t)}{\sqrt{Q_{N_f}(t)}},
\ee
with the polynomial $Q_{N_f}(t)$ depending on the masses of the new
fields, and the same $P_{N_c}(t|h)$ as in (\ref{TCcurve}).
Finally, the 1-differential in all these cases (when the {\it bare}
spectral curve is double-punctured sphere) is
\be
dS_{\rm min} = \frac{t}{i\pi}\frac{dz}{z}.
\ee

\subsection{$SL(N_c)$ Calogero-Moser model}

This time the {\it bare} spectral curve is elliptic $E(\tau)$
\be
L(\xi) = {\vec p}{\vec H} + g\sum_{{\vec \alpha}} F({\vec q}{\vec \alpha}|\xi)
E_{{\vec \alpha}} 
= \left(\begin{array}{ccccc}
     p_1 & gF_{12} & gF_{13} & & gF_{1N_c} \\
     gF_{21} & p_2 & gF_{23} &\ldots & gF_{2N_c} \\
  &&\ldots && \\
     gF_{N_c1} & gF_{N_c2} & gF_{N_c3} && p_{N_c} \end{array}\right).
\label{callax}
\ee
In order to handle the elliptic functions routinely in this case
one needs  more relations than given in the section \ref{Casec}.
Namely,  a few identities for symmetric combinations of $F$-functions are
required \cite{IM}: 
\be
{\cal S}^2(F) = F(q|\xi)F(-q|\xi) = \wp(\xi) - \wp(q), \nn \\
{\cal S}^3(F) = \frac{1}{3}\left(
F(q_{12}|\xi)F(q_{23}|\xi)F(q_{31}|\xi) + {\rm permutations
\ of}\ q_1,q_2,q_3\right) = -\partial_\xi \wp(\xi), \nn \\
\ldots \nn \\
{\cal S}^r(F) = \frac{1}{r}\sum_{\rm perm}
F(q_{12}|\xi)F(q_{23}|\xi)\ldots F(q_{r1}|\xi) = 
\left(-\partial_\xi\right)^{r-2}\wp(\xi).
\ee
With the help of these identities the full spectral curve can be
 represented as
\be
{\cal C}: \ \ \ \ 0 = \det ( t - L(\xi)) =
 \sum_{l=0}^{N_c} S_l(h){\cal T}_{N_c - l}
(t|\xi).
\label{calcurv}
\ee
This time, in varience with the Toda-chain case, eq.(\ref{TCcur}),
the $n$-th order $t$-polynomial ${\cal T}_n(t|\xi)$ - while still
 $h$-independent -
depends  nontrivially on $\xi$:
\be
t^{-n}{\cal T}_n(t|\xi) = 1 \ + \  \nn \\ + 
\sum_{\stackrel{2\leq r_1 < r_2 < \ldots}{m_s > 0}}
\frac{n!}{ \left(n - \sum_s m_sr_s\right)!} 
\prod_s \frac{(-)^{m_s}}{m_s!(r_s!)^{m_s}} 
\left(-\frac{g}{t}\right)^{\sum_s m_sr_s}
\prod_s \left(\partial_\xi^{r_s-2}\wp(\xi)\right)^{m_s}
\label{explT}
\ee
Converting this expression from the flat coordinate $\xi$ to
   the elliptic ones $x$
and $y = \sqrt{\prod_{a=1}^3(x - \hat e_a)}$, one obtains
\be
{\cal T}_0 = 1, \ \ \ {\cal T}_1 = t, \ \ \ {\cal T}_2 = t^2 - x,  \ \ \
{\cal T}_3 = t^3 - 3xt + 2y, \nn \\
{\cal T}_4 = t^4 - 6xt^2 + 8yt - 3x^2 + \sum_{a=1}^3\hat e_a^2, \ \ldots
\ee
- some linear combinations of Donagi-Witten polynomials introduced in
\cite{DW}. An advantage of (\ref{explT}) as compared to \cite{DW} is not only
the simple derivation (and thus the possibility to obtain a general explicit
formula (\ref{explT})), but  the full separation of $h$ and $\xi$
variables achieved in eq.(\ref{calcurv}). See \cite{IM} for details and
discussion.

The 1-differential $dS_{\rm min}$ is, as usual, just
\be
dS_{\rm min} = 2t(\xi)d\xi.
\ee

\subsection{$SL(N_c)$ Ruijsenaars Model}

To finish with our examples, we present a few formulas for the further
generalization of Calogero-Moser system - its relativistic (from one
point of view) or quantum group (from another perspective) generalization:
the $1d$ Ruijsenaars system.
The {\it bare} spectral curve is still elliptic $E(\tau)$, the Lax operator is
composed of the already familiar elliptic functions, but it is given
by a different formula \cite{Ru}:
\be
L_{ij}(\xi) = e^{P_i} \frac{F(q_{ij}|\xi)}{F(q_{ij}|\mu)}
\prod_{l\neq i} n(\mu)\sqrt{\wp(\mu) - \wp(q_{il})}.
\label{RuLax}
\ee
Calogero-Moser model is the $\mu \rightarrow 0$ limit of this one
with the coupling constant $g \sim m$ emerging from the scaling rule
for $P_i$-varibles.
If one takes the normalization function to be $n(\mu) = \sigma(\mu) \sim \mu
+ o(\mu^2)$, rescales as $P_i = \frac{\mu}{g}p_i$, and takes into account that
$\wp(\xi) = \mu^{-2} + o(1)$, $F(q_{ij}|\mu) = 
\mu^{-1}(1 - \delta_{ij}) + \delta_{ij} + o(\mu)$,  it is easy to see that
\be
L_{ij}(\xi) = \delta_{ij}  + \frac{\mu}{g}\left(p_i +
 (1-\delta_{ij})gF(q_{ij}|\xi)
\right) + o (\mu^2),
\ee
and the order-$\mu$ term is exactly the Lax matrix  (\ref{callax}).
For $N_c=2$ the full spectral curve can be written as
\be
0 = \det \left(t\delta_{ij} - L_{ij}(\xi)\right)
= t^2 - t\cdot {\rm tr} L + \det L,
\label{Rucu}
\ee
and 
\be
H \equiv \frac{1}{2n(\mu)}{\rm tr} L = \frac{1}{2}(e^P + e^{-P})
\sqrt{\wp(\mu) - \wp(q)}, \nn \\
\det L = n^2(\mu)\left(\wp(\mu) - \wp(\xi)\right),
\ee
so that  (\ref{Rucu}) gives
\be
t = \frac{H \pm \sqrt{H^2 - \wp(\mu) + \wp(\xi)}}{n(\mu)},
~~{\rm and}~~
n(\mu) dS_{\rm min}^{Ru} = 2n(\mu)td\omega_0  = 2H(\mu)d\omega_0 +
\left.dS_{\rm min}^{Cal}\right|_{h/g^2 = H^2(\mu) - \wp(\mu)}.
\ee
We remind that
\be
dS_{\rm min}^{Cal} \sim \sqrt{\frac{h}{g^2} + \wp(\xi)}d\xi \sim
\frac{\sqrt{\frac{h}{g^2} - x}}{y(x)}dx.
\ee

\sect{Theory of prepotential \label{prepo}}

There are several presentations at this conference devoted to
the prepotential in Seiberg-Witten theory. Instead of repeating
the same things we rather outline here a general theory - not
only applicable to the Seiberg-Witten case (  which is associated
with Riemann surfaces and corresponds to $d=1$ and $\Omega = dS$
below).
This general theory can be given different names:
that of quasiclassical $\tau$-functions, of prepotentials, the
Whitham theory, special geometry etc. - see \cite{Kr2,D,SG,NT,EY}
for various presentations.
Applications to the Seiberg-Witten case are straightforward
see \cite{IM2} and references therein.  
The real meaning of the prepotential theory - and the very fact
that a more fundamental object (prepotential) than the {\it action} exists 
in a rather general setting in classical mechanics - remains obscure.
It should be somehow related to the fundamental role that
{\it quasi}periodic (rather than periodic) trajectories - which exhibit some
ergodicity-like properties -  play in the transition 
from classical to quantum mechanics. Why is the theory of
quasiperiodic trajectories expressible in terms of deformations of
Hodge structures - and how general this statement can be - should
be a subject of further investigation: we do not touch these fundamental
problems in what follows. 

\subsection{Notation and Definitions}

Consider a family ${\cal M}(h)$
of complex  manifolds $M$ of complex dimension $d$ (in the
previous sections $d=1$ and ${\cal M}(h)$ are some families
of spectral {\it curves}, $M = {\cal C}$ ).
 The family is para\-met\-ri\-zed by some {\it moduli} $h_k$,
$k = 1,\ldots,K = {\rm dim}_C{\cal M}$.
Let us fix some canonical system  of $d$-cycles on $M$:
$\ \{A_i, B_i\}$, $i = 1,\ldots,p = \frac{1}{2}{\rm dim} H^{d}(M)$ 
with the intersection matrix $A_i\#B_j = \delta_{ij}$,
$A_i\#A_j = B_i\#B_j = 0$. Finaly, pick up some holomorphic 
$(d,0)$-form $\Omega$ on every $M$.\footnote{
It is clearly
a restriction on $M$ that such $\Omega$ exists:
examples of suitable $M$ are provided by $K3$ ($d=2$) and
Calabi-Yau ($d=3$) manifolds. 
In our discussion below we shall see that this restriction
can sometime be weekend, by admitting $\Omega$'s
with simple singularities.
Additional requirements
for $\Omega$-dependence on moduli  will be specified later.
} 
Its {\it periods}
\be
a_i(h) \equiv \oint_{A_i}\Omega, \ \ \ \ 
a^D_i(h) \equiv \oint_{B_i}\Omega
\label{per}
\ee
are functions of moduli.

Consider now a variation $\delta\Omega$ of $\Omega$ with
the change of parameters (moduli). $\delta\Omega$ is also a 
$(d,0)$-form, not necessarily holomorphic. 
Still, always $\Omega\wedge\delta\Omega = 0$
(just because $\Omega$ is a maximal-rank form), and   integraion of
this relation over entire $M$ gives
\be
0 = \int_M\Omega\wedge\delta\Omega =
\sum_i\left(\oint_{A_i}\Omega\oint_{B_i}\delta\Omega -
\oint_{A_i}\delta\Omega\oint_{B_i}\Omega\right) \ + \ 
{\rm contribution\ from\ singularities}.
\label{origin}
\ee
Imagine that the last item at the r.h.s. - the contribution from
singularities of $\Omega$ and $\delta\Omega$ is absent.
Then we obtain from (\ref{origin}):
\be
\sum_i a_i\delta a^D_i = \sum a^D_i\delta a_i.
\label{symrel}
\ee
This implies that the {\it prepotential}, defined as 
\be
{\cal F} \equiv \frac{1}{2}\sum_i a_ia^D_i = 
\frac{1}{2}\sum_i \oint_{A_i}\Omega\oint_{B_i}\Omega,
\label{Fdef}
\ee
possesses the following property:
\be
\delta{\cal F} = \frac{1}{2}\sum_i\left(a_i\delta a^D_i +
 a^D_i\delta a_i\right)
= \sum_i a^D_i\delta a_i.
\ee
If the freedom of variations is big enough, e.g. if 
$\#K = {\rm dim}_C{\cal M}$
is the same as $\#p = \frac{1}{2}{\rm dim}\ H^{d}(M)$,
we conclude from this that
\be
a^D_i = \frac{\partial{\cal F}}{\partial a_i}
\label{aDfroF"}
\ee
and
\be
{\cal F} = \frac{1}{2}\sum_i a_ia^D_i =
\frac{1}{2}\sum_i a_i\frac{\partial {\cal F}}{\partial a_i}.
\label{homF"}
\ee
In other words, we can consider $a_i$ as independent variables,
and introduce the prepotential ${\cal F}(a)$ by the rule (\ref{Fdef}) -
and it will always be a homogeneous function of degree 2 -
as follows from (\ref{homF"}).

The two  requirements built into this simple construction are

(i) the absence of singularity contributions at the r.h.s. of (\ref{origin});

(ii) the matching between the quantities of moduli and $A$-cycles,
$$
K \equiv {\rm dim}_C {\cal M} =  p \equiv \frac{1}{2}{\rm dim} H^{d}(M) 
$$.

\subsection{Comments on requirement (i)}

The problem with this restriction is that variation of holomorphic
object w.r.to moduli usually makes it singular - by the very definition
of moduli of complex structure. Thus, even if $\Omega$ is free of
singularities one should expect them to appear in $\delta\Omega$.
The only way out would be to get the newly emerging poles
cancelled by zeroes of $\Omega$ - but often the space of holomorphic
$\Omega$'s is too small to allow for adequate adjustement.

Fortunately, requirement (i) can be made less restrictive. 
One can allow to consider $\Omega$ which is not holomorphic,
but possesses {\it simple} singularities at isolated divisors. As 
a pay for this it is enough to  enlarge the set of $A$-cycles, by adding
the ones wrapping around the singularity divisors, and also add all 
independent $B$-chains, connecting these divisors (such that
$\partial B = div_1 - div_2$).\footnote{
For example, if $M$ is a complex curve ($d=1$), $\Omega$ can be
a meromorphic $(1,0)$-differential with {\it simple} (order one) poles
at some punctures $\xi_\alpha$, $\alpha = 0,1,\ldots,r$. Then one
should add $r$ circles around the points $\xi_1,\ldots,\xi_r$ to the
set of $A$-cycles, and $r$ lines (cuts) connecting $\xi_0$
with $\xi_1,\ldots,\xi_r$ to the set of $B$-contours in eq.(\ref{origin}).
Then the last term at the r.h.s. can be omitted in exchange for enlarging
the sum in the first term.
} 
At the same time residues at the simple singularities should be added to 
the set of moduli $\{h\}$, thus preserving the status of the
second requirement (ii).
This prescription is still not complete, because the integrals over
newly-added $B$-chains are divergent (because these end
at the singularities of $\Omega$). However, the structure of divergence  
is very simple: if a cut-off is introduced, the cut-off-\-dependent
piece in ${\cal F}$ is exactly quadratic in the new moduli -
and does not depend on the old ones. If one agrees to
define the prepotential - which is generic homogeneous
function of order two - modulo {\it quadratic} functions of
moduli, the problem is resolved.
Thus the real meaning of constraint (i) is that $\delta\Omega$ should
not introduce new singularities as compared to $\Omega$ - so that we
do not need to introduce new cycles, thus new moduli, derivatives
over which would provide new singularities.
Since now the freedom to choose $\Omega$ is big enough, such
special adjustement is usually available.

The non-simple singularities (higher-order poles at divisors)
should be resolved - i.e. considered as a limit of several simple
ones when the corresponding divisors tend to coincide. The
corresponding $B$-chains shrink to zero in the limit, but integrals
of $\Omega$ over them do not vanish, if $\Omega$ is indeed
singular enough. This procedure of course depends on a particular
way to resolve the non-simple singularity. Essentially, if we
want to allow  the one of  an arbitrary type on the given divisor,
it is necessary to introduce  coordinate system in the vicinity
of  the divisor and consider all the negative terms of Laurent
expansion of $\Omega$ as moduli, and ''weighted'' integrals
around the divisor as $A$-cycles. In the case of $d=1$, when
the divisors are just points (punctures) one can easily
recognize in this picture the definition of KP/Toda-induced
Whitham prepotential 
with one-parameter set of ''time''-variables
(Laurent expansion coefficients or moduli of coordinate
systems) for every puncture as
additional moduli (see, for example, \cite{NT,IM2} and references therein). 
As often happens, it is most natural from the
point of view of string theory (integrability theory in this case)
to put all the moduli in a single point (or two), but from the
point of view of algebraic geometry it is better to redistribute
them as simple singularities at infinitely many divisors. 

Finally, singularities of $\Omega$ on subspaces of codimension higher
than one do not contribute to eq.(\ref{origin}) at all  - and often
variation w.r.to moduli produces only singularities of such type 
as $d>1$.

\subsection{Requirement (ii)}

Thus, what essentially remains is the other requirement (ii) -
the matching condition between the number of moduli and $A$-cycles.
Since the procedures involved in resolution of (i) do not change this
matching (they always add as many new moduli as new $A$-cycles),
this requirement can be analyzed at the very beginning -
before even introducing $\Omega$.

\sect{Picard-Fuchs equations}

Dependence $a^i(h)$, $a_i^D(h)$, described by eq.(\ref{per}) can
be also formulated in terms of differential Picard-Fuchs equations
for the cohomology classes of $\Omega$. They are often
convenient for comparison of Whitham universality classes of 
different models: Whitham-equivalent models should have 
equivalent Picard-Fuchs equations.  We refer to \cite{IM2} on
details of how this idea can be elaborated on. Here we just list
some important examples (all for $G=SL(2)$).

\subsection{Pure gauge $N=2$ SUSY model in $4d$: sine-Gordon
model in $1d$}
In this case
\be
dS_{\rm min} \sim \sqrt{\frac{\tilde x - u}{\tilde x^2 - \Lambda^4}}d\tilde x
\ee
and Picard-Fuchs equation is \cite{Pic}
\be
\left(\frac{\partial^2}{\partial u^2} + \frac{1}{4(u^2-\Lambda^4)}\right)
\oint dS_{\rm min} = 0.
\label{PFsG}
\ee

\subsection{The flow from pure gauge $N=4$ SUSY model to the
$N=2$ SUSY one in $4d$: Calogero-Moser model in $1d$}
Now
\be
dS_{\rm min} \sim \sqrt{\frac{\hat h - x}{(x-\hat e_1)(x - \hat e_2)
(x - \hat e_3)}}dx.
\label{dSCal}
\ee
The branching points $\hat e_a$ of the elliptic {\it bare} spectral
curve $E(\tau)$ are functions of $\tau$, 
and the simplest Picard-Fuchs equation looks like
\be
\frac{1}{2\pi i}\frac{\partial}{\partial\tau}\oint dS_{\rm min} = 
\left(y^2(\hat h)\frac{\partial^2}{\partial\hat h^2} +
\left[\frac{1}{2}\hat h^2 - \frac{1}{2}\hat hC(\tau) - \frac{1}{12}
{\rm g}_2(\tau)\right]\frac{\partial}{\partial\hat h}\right) .
\oint dS_{\rm min}
\label{PFCal}
\ee
Here $C(\tau)$ is just the same quasimodular form 
(  the logarithmic derivative of  the Dedekind function) that appeared in 
(\ref{Cdef}) above, while the modular form 
${\rm g}_2(\tau) = \frac{2}{3}\left[\theta_{00}^4(\tau) +
\theta_{01}^4(\tau) + \theta_{10}^4(\tau)\right]$.
The differential operator
eq.(\ref{PFCal}) is presumably convertible (by conjugation and
 change of variables) to the Schroedinger form:
\be 
\frac{1}{2\pi i}\frac{\partial}{\partial \tau} -
\left(y(\hat h)\frac{\partial}{\partial\hat h}\right)^2 - \hat h =
 \frac{1}{2\pi i}\frac{\partial}{\partial \tau} -
\frac{\partial^2}{\partial\chi^2} - \wp(\chi),
\ee 
which (if true) would reflect  the interpretation of Whitham theory as
that of quantization: (we consider essentially classical Calogero-Moser
model, but the Picard-Fuchs equation on the moduli space is
the Schroedinger equation for this model). See \cite{IM2} for more
details.

One can derive  an infinite set of Picard-Fuchs equations, with
different powers of $\tau$-derivative, eq.(\ref{PFCal}) being the
simplest one in the series (first $\tau$-derivative). Only two of them are
algebraically independent, because $\oint dS_{\rm min}$ depends
only on two variables: $\hat h$ and $\tau$. Still, the entire infinite
series, once derived, can exhibit some new nice structure - as it
usually happens (compare with the Virasoro etc constraints in
matrix models).

\subsection{Ruijsenaars model in $1d$}
\be
dS^{Ru}_{\rm min} = H(\mu)\frac{dx}{\sqrt{(x-\hat e_1)(x - \hat e_2)
(x - \hat e_3)}} + \left.dS^{Cal}_{\rm min}\right|_{\hat h = H(\mu)^2 -
\wp(\mu)}.
\ee
Here $dS^{Cal}_{\rm min}$ is given by (\ref{dSCal}).
Picard-Fuchs equation is not drastically different from (\ref{PFCal}),
most important, the lowest equation seems to be still of the first order
in $\partial/\partial\tau$ - what does not allow to identify it with the
Picard-Fuchs equation for Calabi-Yau model, where all the
derivatives are of the second order  (see \cite{IM2} and
below). Again, there are only two independent Picard-Fuchs
equations.

\subsection{The $WP^{12}_{1,1,2,2,6}$-induced Calabi-Yau
model}
The manifold is a factor of the one, defined by the equation
\be
0 = p(z) = \frac{z_1^{12}}{12} + \frac{z_2^{12}}{12} +
\frac{z_3^6}{6} + \frac{z_4^6}{6} + \frac{z_5^2}{2} +
\phi\frac{z_1^6z_2^6}{6} + \psi z_1z_2z_3z_4z_5
\label{CYeq}
\ee
The 3-form $\Omega$, which is used in the construction of
the prepotential on the lines of  s.\ref{prepo}, is a restriction
of $dz_1\wedge dz_2 \wedge dz_3\wedge dz_4\wedge dz_5$
(one should take into account the quasihomogeneity of
(\ref{CYeq}) - this allows to eliminate {\it two} variables to
get a 3-form). Thus its periods are proportional to
\be
\oint \Omega \sim \int {\cal D}\lambda(z) \int dz_1\ldots dz_5
e^{i\lambda(z)p(z)} \sim 
\int dz_1\ldots dz_5 
\left(\int d\lambda e^{i\lambda p(z)}\right)  \sim \nn \\
\sim \int dz_1\ldots dz_5 e^{p(z)}
\label{CYper}
\ee
(the quasihomogeneity of $p(z)$ is used to eliminate $\lambda$)
and satisfy the set of Picard-Fuchs equations (which are nothing
but Ward identities for the integral (\ref{CYper})). The simplest one
is
\be
\left[\left(\frac{\partial}{\partial\phi}\right)^2 -
\left(\phi\frac{\partial}{\partial\phi} + \frac{1}{6}\psi\frac{\partial}
{\partial\psi} + \frac{1}{6}\right)^2\right] \oint\Omega = 0.
\label{PFCY}
\ee
Again, since there are two moduli, only two equations from the whole
set will be algebraically independent.

In the particular (''conifold'') double scaling limit, when $\phi, \psi
\rightarrow \infty$ with $\frac{u}{\Lambda^2} = \phi - i\psi^6$
fixed, this equation (\ref{PFCY}) reduces exactly to
the sine-Gordon-case one, eq.(\ref{PFsG}). This reflects the fact
that in the target space this limit cortresponds to the
$\alpha' \rightarrow 0$ limit, when the $d=10$ Calabi-Yau model
reduces to the $d=4$ one - and  the sector described by the
periods of $\Omega$ is exactly the gauge sector described by
the Seiberg-Witten theory. See \cite{KKLMV} for details and
references.

There is no doubt that the equation (\ref{PFCY}) itself, not only
its conifold limit, can be represented in terms of some simple
$1d$ system. However, at the moment we do not know what this
system is. Neither Calogero nor Ruijsenaars models seems to suit.
Technically, the lowest Picard-Fuchs equations for these models
contain only first derivative w.r.t. one of the variables ($\tau$), while
in (\ref{PFCY}) both derivatives are of the second order. Physically,
the relevant models should not be assocaited with particular
groups (only the rank of the group should be fixed): this is
because the variation of moduli of Calabi-Yau model can change
one group for another (in one point of the moduli space one can 
have $SL(3)$ symmetry, while in another one it would be
$SL(2)\times SL(2)$ - and neither one is a subgroup of another).
This phenomenon is not {\it directly} relevant for the rank-one
example of eq.(\ref{PFCY}) - still it explains why Calogero model
itself should not be enough - and shows the direction for the
search of the relevant models.

\sect{Instead of conclusion}

Following refs.\cite{Go,IM,IM2} we presented some evidence
that the results like those of \cite{SW,SW2} can be nicely
systematized in the {\it a priori} different language - that of
the $1d$ integrable systems. We do not find it very surprising,
because the question that was addressed in \cite{SW,SW2}
is very special: the one about the {\it low-energy} effective
actions, and the adequate terms in which the conformally
invariant theories in the deep UV are related to the topological ones
in the deep IR are necessarily rather simple.

In fact, the general scheme that one can keep in mind is as follows
\cite{UFN}:
exact Wilsonian effective actions, defined by the functional
integrals like
\be
e^{S_{\rm eff}(t|\Phi)} = \int_\Phi {\cal D}\phi\  e^{S_t(\phi)} 
\ee
naturally depend on two kinds of variables: the coupling constants
$t_{k_x}$ in the bare action
$S_t(\phi) \sim$ $\sum_{\{k_x\}} t_{\{k_x\}} {\rm tr} \prod_x
\phi_x^{k_x}$
and the background fields $\Phi$ (examples of the latter
ones are our $a^i$ and $a_i^D$ above). Such exact effective
actions (the generating functionals for all the correlators in the
given field theory)
- as one knows well from the example of matrix models -
are ''infinitely symmetric''  because of the freedom to change
integration variables. This symmetry is often enough to
identify them with pure algebraic objects: generalized
$\tau$-functions, defined as generating functions of all the
matrix elements of a universal group element $g$,
$\tau(t|g) = \sum_{k_x,\bar k_x} \langle k_x|g|\bar k_x\rangle
t_{\{k_x,\bar k_x\}}$. In general case (non-vanishing normalization
point) both effective action and the $\tau$-function (for quantum
group) are operator-valued; the IR stable point of RG flow
should then correspond to the classical limit in the group
theory language.

Such considerations are, of course, very
general and can seem almost senseless: still they imply something
both in the general framework (for example, so defined $\tau$-functions
always satisfy some bilinear Hirota-like equations), and in concrete
examples. The most famous example is the one
of matrix models. Another - newly emerging example - is that
of the low-energy theories: restricting consideration to the IR
stable points of renormalization group flow, one drastically 
diminishes the number of degrees of freedom (moduli) - what in the
group-theory language corresponds to consideration of small
enough groups (not necessarily the 3-loop group, as at generic
normalization point for the $4d$ field theory).\footnote{
A nice particular example of  the relation between RG flows and
integrability theory is by now famous identity
$$
\beta_W\langle {\rm tr}\ \phi^2\rangle \ \sim \  
2{\cal F}_{\rm red} - \sum_i a^i\frac{\partial{\cal F}_{\rm red}}
{\partial a^i},
$$
- a member of the anomaly family (together with
$\beta_W\langle{\rm tr}\ G^2 \rangle \sim T_{\mu\mu}$ and
axial anomaly), where the l.h.s. is clearly of RG nature and
the r.h.s. represents the breakdown of homogeneity of 
the prepotential ${\cal F}$   which occurs by fixing
 one of its arguments (the scale $\Lambda$).
} 

To put it differently,
various theories flow to the same universality class in the IR
limit - thus these classes can be (and are) rather simple. 
What the general identification of effective actions with the
tau-functions (i.e. with group theory) teaches us is that these
classes should be also representable by some $\tau$-functions.
However, these cannot be just conventional $\tau$-functions -
defined in the Lie-group terms - because some parameter of the
effective action (the normalization point) is fixed. But in
order to understand {\it what} are these relevant objects one
can consider just the RG flow within some simple enough
integrable system - and then discover that the relevant objects
are {\it quasiclassical} $\tau$-functions (or prepotentials).
This can provide a kind of an explanation of why it was natural
to {\it try} to identify the results of \cite{SW,SW2} with those
of integrability theory and where exactly (the Whitham theory) 
one had to look for this identification. This also explains why there
should be no big surprise once such correspondence is
established.

What needs to be understood, however, is the general description
of how  group theory (represented by generalized $\tau$-functions)
always flows to that of Hodge deformations (represented by 
prepotentials). The main message of this presentation can be
that such phenomenon  exists, and one should think
of what could be the reasons behind this and what is the
adequate technical approach to a more generic situation. Once found,
the answers can
shed new light on the implications of symmetries (group theory)
for the low-energy dynamics and algebraic geometry (of
moduli spaces) - and this would be of definite use for the
future developement of quantum field and string theory.

\bigskip

\bigskip

{\Large{\bf Acknowledgements}}

\bigskip

We are indebted for discussions to
O.Aharony,  E.Akhmedov, L.Alvarez-\-Gaume, S.Das, A.Gera\-simov, 
C.Gomez, A.Gorsky, A.Hanany, S.Kachru, 
I.Kri\-che\-ver, H.Kunitomo, A.Losev, A.Mar\-shakov, A.Mi\-ro\-nov, 
T.Nakatsu, N.Ne\-kra\-sov,
K.Oh\-ta, M.Ol\-sha\-netsky, 
A.Rosly, V.Rub\-tsov, J.Son\-nen\-schein, A.Stoyanovsky,
K.Takasaki and A.Tokura.  

H.I. is supported in part by Grant-in-Aid for Scientific Research
(07640403) from the Ministry of Education, Science and Culture, Japan.
A.M. acknowledges the
hospitality of  Osaka University and  the support of the JSPS. 
A.M. also thanks the organizers for a unique chance to participate
in this very stimulating meeting.

\end{document}